\documentclass[epj]{svjour}
\usepackage{graphics}
\usepackage{color}
\usepackage{graphicx}
\usepackage{epstopdf}
\usepackage{subfigure}
\usepackage{amssymb}
\usepackage{CJK}
\usepackage{indentfirst}
\usepackage{amsmath}
\usepackage[square, comma, sort&compress, numbers]{natbib}
\usepackage[table]{xcolor}

\begin{document}

\title{Model-independent constraints on cosmic curvature: implication from the future gravitational wave observation DECIGO}

\author{Xiaogang Zheng\inst{1}, Shuo Cao\inst{2}\thanks{\emph{e-mail:} caoshuo@bnu.edu.cn}, Yuting Liu\inst{2}, Marek Biesiada\inst{2,3}, Tonghua Liu\inst{2,3}, Shuaibo Geng\inst{2}, Yujie Lian\inst{2}, Wuzheng Guo,\inst{2}}

\institute{School of Electrical and Electronic Engineering, Wuhan Polytechnic University, Wuhan 430023, China;
\and Department of Astronomy, Beijing Normal University, Beijing, 100875, China;
\and National Centre for Nuclear Research, Pasteura 7, 02-093 Warsaw, Poland}

\date{Received: date / Revised version: date}

\abstract{
A model-independent test of the cosmic cosmic curvature parameter
$\Omega_k$ is very important in cosmology. In order to estimate
cosmic curvature from cosmological probes like standard candles, one
has to be able to measure the luminosity distance $D_L(z)$, its
derivative with respect to redshift $D'_L(z)$ and independently know
the expansion rate $H(z)$ at the same redshift. In this paper, we
study how such idea could be implemented with future generation of 
space-based DECi-hertz Interferometer 
Gravitational-wave Observatory (DECIGO), in combination with cosmic chronometers
providing cosmology-independent $H(z)$ data. Our results show that
for the Hubble diagram of simulated DECIGO data acting as a new type
of standard siren, it would be able to constrain cosmic curvature
with the precision of $\Delta \Omega_k= 0.09$ with the currently
available sample of 31 measurements of Hubble parameters. In the
framework of the third generation ground-based gravitational wave
detectors, the spatial curvature is constrained to be $\Delta
\Omega_k= 0.13$ for Einstein Telescope (ET). More interestingly,
compared to other approaches aiming for model-independent
estimations of spatial curvature, our analysis also achieve the
reconstruction of the evolution of $\Omega_k(z)$, in the framework
of a model-independent method of Gaussian processes (GP) without
assuming a specific form. Therefore, one can expect that the newly
emerged gravitational wave astronomy can become useful in local
measurements of cosmic curvature using distant sources. }
\PACS{{cosmological}{parameters}
\and {cosmology}{observations}}

\authorrunning{Xiaogang Zheng, et al.}

\titlerunning{Model-independent constriants on $\Omega_k$ from future DECIGO}
\maketitle

\section{Introduction}\label{sec:introduction}
The spatial curvature parameter $\Omega_k$  is a very significant
quantity closely related to many fundamental issues in modern
cosmology, such as the structure and evolution of the Universe
\citep{Cao2019a,Qi2019a}. The most popular  concept of the very
early Universe undergoing an exponential phase of expansion predicts
that the radius of curvature of the Universe should be very large,
which means that cosmic curvature should be close to zero
\citep{Weinberg2013}. Current cosmological observations, e.g., the
combined \textit{Planck} 2018 cosmic microwave background (CMB) and
baryon acoustic oscillation (BAO) measurements, strongly favor this
inflation theory and demonstrate the flatness of the Universe
($\Omega_k=0.001\pm0.002$) \citep{Planck Collaboration2018}.
However, one should note that such stringent constraint on the
cosmic curvature is indirect and strongly relying on the
pre-assumption of a specific cosmological model (i.e., the
cosmological constant plus cold dark matter model, usually
abbreviated as $\Lambda$CDM model). In general, most studies
focusing on the cosmic curvature always assume that dark energy is
just a cosmological constant, while  the Universe is assumed flat in
most of the dark energy studies. However,  recent analysis indicated
that cosmological constant assumption might cause tension between
$\Lambda$CDM and dynamical dark-energy model, while flat Universe
assumption may lead to an incorrect reconstruction of the dark
energy equation of state
\citep{Ichikawa2006,Clarkson2007,Gong2007,Virey2008,Li2018,Cao2019b}.
Besides, the combination of the \textit{Planck} 2018 $TT,TE,EE+lowE$
power spectra data alone slightly favor a mildly closed Universe,
i.e., $\Omega_k= -0.044^{+0.018}_{-0.015}$ \citep{Planck
Collaboration2018,Valentino2019}. Any small change in the spatial
curvature could  have a huge impact on the reconstructed history of
the Universe. Therefore, purely geometrical and model-independent
methods of inferring the spatial curvature have always been an
important issue in cosmology.

In particular, the distance sum rule \citep{Bernstein2006}, which
characterizes the relation between the distances of the background
source, the lens and the observer in the
Friedmann--Lema\^{\i}tre--Robertson--Walker (FLRW) metric, has been
proposed as a model-independent method to constrain the curvature of
the Universe \citep{Rasanen2015}. Such methodology was then applied
to test the validity to FLRW metric, based on the galactic-scale
lensing systems where strongly lensed gravitational waves and their
electromagnetic counterparts can be simultaneously detected
\citep{Cao2019b}. More recently, \citet{Qi2019b,Zhou2020} extended
the cosmic curvature analysis to higher redshift, using the latest
data sets of strong lensing systems \citep{Cao2015,Chen2019}
combined with intermediate-luminosity quasars calibrated as standard
rulers \citep{Cao2017a}. Another straightforward method to constrain
the cosmic curvature has been proposed by \citet{Clarkson2007},
using the expansion rate measurements H(z) and the transverse
comoving distances $D(z)$. Such method is also model-independent,
which has been further developed with updated observational SNe Ia
data acing as standard candles
\citep{Shafieloo2010,Mortsell2011,Sapone2014,Cai2016,Wang2019,Li2016c,Wei2017,Wang2017,Rana2017},
ultra-compact structures in radio quasars acting as standard rulers
\citep{Cao2019b}, and the Hubble diagram based on the non-linear
relation between the UV and X-ray monochromatic luminosities of
quasars \citep{Risaliti2018,Melia2019,Wei2020}.

As a new window on the Universe, gravitational wave (GW) signals
create alternative opportunities. Namely, the GW signals from
inspiralling  binary black holes (BH-BH) and neutron stars (NS-NS)
(or mixed BH-NS systems) can be used as standard sirens providing
the luminosity distances in a direct way, not relying on the cosmic
distance ladder \citep{Schutz1986}. Compared with the observations
of SNe Ia in the electromagnetic (EM) domain, the greatest advantage
of GWs is the independent calibration of luminosity distances. Some
recent studies have discussed the possibility of extending the
cosmic curvature test based on the simulated data of GW from future
gravitational wave detectors \citep{Jimenez2018,Wei2018,Liao2019}.
In this paper, we will consider the reconstruction of cosmic
curvature parameter $\Omega_k$ at different redshifts
\citep{Clarkson2007}, focusing on the standard sirens accessible to
the future space-borne GW detector, i.e., the DECi-hertz
Interferometer Gravitational wave Observatory (DECIGO), as well as
the third generation GW ground-based detector, i.e., the Einstein
Telescope (ET).  In the EM domain, the recent measurements of Hubble
parameters $H(z)$ are inferred from the differential ages of
galaxies, i.e., cosmic chronometers (CC) acting as standard clocks.
In order to investigate the redshift dependence of $\Omega_k$
without assuming any specific functional form, the so-called
Gaussian processes (GP) \citep{Seikel2012a} will also be applied to
reconstruct the evolution of the curvature of the universe, which
has been widely used in recent works
\citep{Yang2015,Yu2016,Liu2019,Wu2020,Zheng2020}. This paper is
organized as follows. In Sec. 2, we give a brief introduction of the
methodology and data used in this work. Our results and discussions
are presented in Sect.~3. Finally, the general conclusions are
summarized in Sect.~4.

\section{Methodology and observational data}

Under the assumption that the Universe is homogeneous and isotropic
on the large scales, the FLRW metric can be used to describe its
geometry as
\begin{equation}
ds^2 = -dt^2 + a^2(t)[\frac{dr^2}{1-Kr^2}+r^2(d\theta^2+sin^2\theta
d\phi^2)],
\end{equation}
where $t$ is the cosmic time and ($r$,$\theta$,$\phi$) are the
comoving spatial coordinates. The scale factor $a(t)$ is the only
gravitational degree of freedom and its evolution (by virtue of the
Einstein's equations) is determined by the matter and energy of the
Universe. The dimensionless curvature $K=-1,0,+1$ corresponds to
open, flat and closed Universe, respectively. In such metric, the
luminosity distance $D_L(z)$ can be expressed as
\begin{equation}\label{eq2}
D_L(z) = \left\lbrace \begin{array}{lll}
\frac{c(1+z)}{H_0\sqrt{|\Omega_{\rm k}|}}\sinh\left[\sqrt{|\Omega_{\rm k}|}\int_{0}^{z}\frac{dz'}{E(z')}\right]~~{\rm for}~~\Omega_{K}>0,\\
\frac{c(1+z)}{H_0}\int_{0}^{z}\frac{dz'}{E(z')}~~~~~~~~~~~~~~~~~~~~~~~{\rm for}~~\Omega_{K}=0, \\
\frac{c(1+z)}{H_0\sqrt{|\Omega_{\rm k}|}}\sin\left[\sqrt{|\Omega_{\rm k}|}\int_{0}^{z}\frac{dz'}{E(z')}\right]~~~~{\rm for}~~\Omega_{K}<0.\\
\end{array} \right.
\end{equation}
The dimensionless Hubble parameter $E(z)$ is defined as $H(z)/H_0$,
where $H(z)$ is the Universe expansion rate and $H_0$ is the Hubble
constant. The curvature parameter $\Omega_k$ is related to $K$ as
$\Omega_k=-c^2 K/(a_0H_0)^2$, where $c$ is the speed of light. The
derivative of Eq.~(2) will provide the cosmic curvature $\Omega_k$,
expressed by the Universe expansion rate $H(z)$ and transverse
comoving distance $D(z)$ as  \citep{Clarkson2007}
\begin{equation}\label{eq3}
\Omega_k=\frac{[H(z)D^{'}(z)]^2-c^2}{[H_0D(z)]^2}.
\end{equation}
The luminosity distance $D_L(z)$ is simply related to the transverse
comoving distance $D(z)$ as $D(z)=D_L(z)/(1+z)$ \citep{Hogg1999},
while $D^{'}(z)=dD(z)/dz$ denotes the derivative with respect to
redshift $z$. In our analysis, we use the luminosity distance as a
function of redshift $z$ based on simulated standard siren data from
DECIGO and ET to reconstruct $D(z)$ and $D^{'}(z)$, independently,
and combine these two reconstructions with independent $H(z)$
measurements to derive $\Omega_k$.

\subsection{Distance from GW standard siren DECIGO and ET}

DECi-hertz Interferometer Gravitational-wave Observatory (DECIGO)
project proposed by Japan \citep{Kawamura2011,Seto2011} is a future
space gravitational-wave antenna, whose currently most attractive
science objective is detecting the gravitational waves from the
inflation \citep{Kawamura2019b}. Unlike the ground-based GW
detectors (such as aLIGO and VIRGO), DECIGO was designed most
sensitive between 0.1 Hz and 10 Hz to reach lower frequency
detection with drag-free spacecrafts. More importantly, this
frequency range fills the gap between sensitivity windows of ground
based detectors and  future LISA space-borne detector. In
particular, this opportunity would allow the early detection (and
measurements) of inspiralling sources which would enter the
ground-based detector a few years after. This would create
unprecedented opportunity to  precisely measure properties of such
sources and to improve the determination of their position on the
sky. Aiming to improve the detection sensitivity, DECIGO implements
four clusters of spacecraft, and each cluster consists of three
spacecraft with three Fabry-Perot Michelson interferometers, whose
arm length is 1000 km. The expected sensitivity is $10^{-25}$
Hz$^{-1/2}$ for two clusters of DECIGO at the same position for
three years of mission \citep{Kawamura2019a}.
Without any pre-assumptions concerning cosmology, signals from
inspiraling binary neutron stars (NS-NS) and black holes (BH-BH) can
provide an absolute measurement of the luminosity distance
\citep{Abbott2016,Abbott2017}. The GW amplitude is related to the
so-called chirp mass (which can be measured from the GW signal's
phasing) and the luminosity distance. Therefore, the luminosity
distance can be extracted from the amplitude and the rate of
frequency change. DECIGO will be able to detect the gravitational
waves from neutron star binaries even at a redshift of $z\sim5$ for
five years of its mission and many intermediate-mass black hole
binary coalescences.

Focusing on the GW signals from the binary system with component
masses $m_1$ and $m_2$, one can define the chrip mass
$M_c=M\eta^{3/5} $, where $M$ is the total mass of binary system
$M=m_1+m_2$ and $\eta$ is symmetric mass ratio $\eta= m_1m_2/M^2$.
In GW experiments, the Fourier transform of time domain waveform can
be computed as
\begin{equation}\label{eq4}
\widetilde{h}(f)=\frac{A}{D_L(z)}M^{5/6}_zf^{-7/6}e^{i\Phi(f)},
\end{equation}
where $A$ is a geometrical average over the inclination angle of the
binary system  $A=(\sqrt6\pi^{2/3})^{-1}$, $M_z$ -- the so called
redshifted chirp mass, is defined as $M_z=(1+z)M_c$ and $D_L(z)$ is
the luminosity distance. $\Phi(f)$ is the frequency-dependent phase
caused by orbital evolution which is usually described by
post-Newtonian (PN) approximation (an approximation to General
Relativity in the weak-field, slow-motion regime) of different order
\citep{Kidder1993,Cutler1994}.

In order to estimate the uncertainty of the measurement of the
luminosity distance, one can use the following Fisher matrix
\begin{equation}\label{eq5}
\Gamma_{ab}=4Re\int_{f_{min}}^{f_{max}}\frac{\partial_a\widetilde{h}^{*}_i(f)\partial_b\widetilde{h}_i(f)}{S_h(f)}df,
\end{equation}
where $\partial_a$ ($\partial_b$) is derivative with respect to
parameter $\theta_a$ ($\theta_b$). The noise power spectrum $S_h(f)$
consists of the shot noise, the radiation pressure noise and the
acceleration noise. Based on the mechanical parameters of the DECIGO
tuned to the $0.1-10\;Hz$ frequency window, i.e., the arm length
1000km, the output laser power 10W with wavelength $\lambda$=532nm,
the mirror diameter 1m with its mass 100 kg, and the finesse of
Fabry-Perot Michelson interferometer cavity 10, the noise power
spectrum is fitted as \citep{Kawamura2006}
\begin{eqnarray}
S_h(f)=6.53\times10^{-49} \left[ 1+\left(\frac{f}{7.36Hz}\right)^{2} \right] \nonumber \\
+4.45\times10^{-51}\times \left(\frac{f}{1Hz}\right)^{-4}\times \frac{1}{1+\left(\frac{f}{7.36Hz}\right)^{2}}  \nonumber \\
+4.94\times10^{-52}\times \left(\frac{f}{1Hz}\right)^{-4}
\rm{Hz^{-1}} .
\end{eqnarray}
For the convenience of calculations, we assume equal mass NS-NS
binary system with $m_1=m_2 = 1.4\;M_\odot$. Then we have $M_z =
1.22(1 + z)M_\odot$ and $\eta=0.25$. Therefore, the luminosity
distance is independent on other parameters in the Fisher matrix and
the instrumental uncertainty of the measurement of the luminosity
distance can be estimated as
\begin{equation}\label{eq7}
\sigma^{instr}_{D_L}=\sqrt{\Gamma_{aa}^{-1}}.
\end{equation}
Concerning the uncertainty budget, the luminosity distance precision
per GW is taken as \citep{Cai2017}
\begin{equation}\label{eq8}
\sigma_{D_L}=\sqrt{(\sigma^{instr}_{D_L})^2+(\sigma^{lens}_{D_L})^2}.
\end{equation}
In our simplified case, $\sigma^{instr}_{D_L}\simeq
\frac{2D_L}{\rho}$, where $\rho$ denotes the signal-to-noise ratio
(SNR) of DECIGO interferometers \citep{Zhao2011}. Meanwhile, the
lensing uncertainty caused by the weak lensing can be estimated as
$\sigma^{lens}_{D_L}=0.05zD_L$ \citep{Sathyaprakash2010}. As a
result, the total uncertainty of $D_L$ is modeled as
\begin{eqnarray}
\sigma_{D_L}=\sqrt{\left( \frac{2D_L}{\rho} \right)^2+(0.05zD_L)^2}
\end{eqnarray}

Finally, we employ the redshift distribution of the GW sources
observed on Earth expressed as \citep{Sathyaprakash2010}
\begin{equation}\label{eq9}
P(z)\propto \frac{4\pi D^2_c(z)R(z)}{H(z)(1+z)},
\end{equation}
where $D_c(z)$ is comoving distance, $H(z)$ the expansion rate.
$R(z)$ is the NS-NS coalescence rate, which according to
\citep{Schneider2001,Cutler2009} can be approximated as:
\begin{equation}\label{eq10}
R(z) = \left\lbrace \begin{array}{lll}
1+2z,~~~~~z\le1\\
\frac{3}{4}(5-z),~~1<z<5 \\
0,~~~~~~~~~~~~z\ge5.\\
\end{array} \right.
\end{equation}
In our simulation, we assume the flat $\Lambda$CDM as our fiducial
cosmological model with the matter density parameter and the Hubble
constant values ($\Omega_m=0.315$, $H_0=67.4\;km\;s^{-1}Mpc^{-1}$)
taken after {\it Planck} CMB measurements \citep{Planck
Collaboration2018}. Recent analysis of \citep{Kawamura2019a}
suggests that the space-based GW detector DECIGO can detect up to
10,000 GW events up to redshift $z\sim5$ in one year of operation.
Thus, we simulate a mock data of 10,000 GW events to be used for the
cosmic curvature constraint analysis. The redshift distribution of
our mock catalog is shown in Fig.~1. In order to improve the
reconstruction of $D(z)$ we also simulate GW events observable by
the Einstein Telescope (ET), which is the planned third-generation
ground-based GW detector \citep{ET2015}. Compared to the current
advanced ground-based detectors, ET was designed more sensitive
between 1 Hz and $10^4$ Hz. This and increased overall sensitivity,
will expand the detection space by three orders of magnitude
\citep{ET2015}. As proposed by the design document, it is made up of
three collocated underground detectors, each with 10 km arm and with
a 60 degree opening angle. Theoretically, ET will be able to detect
GWs  from the NS-NS mergers up to the redshift $z\sim2$ and BH-NS
mergers up to the redshift of $z\sim5$ \citep{Cai2017}. In our
analysis, we perform the Monte Carlo simulation of ET detectable GW
signals from NS-NS systems up to $z\sim2$. Specific simulation steps
are similar to those used in \citep{Qi2019c,Qi2019d}. The mock
catalog of 1000 simulated GW events observable by the ET is
displayed in Fig.~2.

\begin{figure}
\begin{center}
\includegraphics[width=0.95\linewidth]{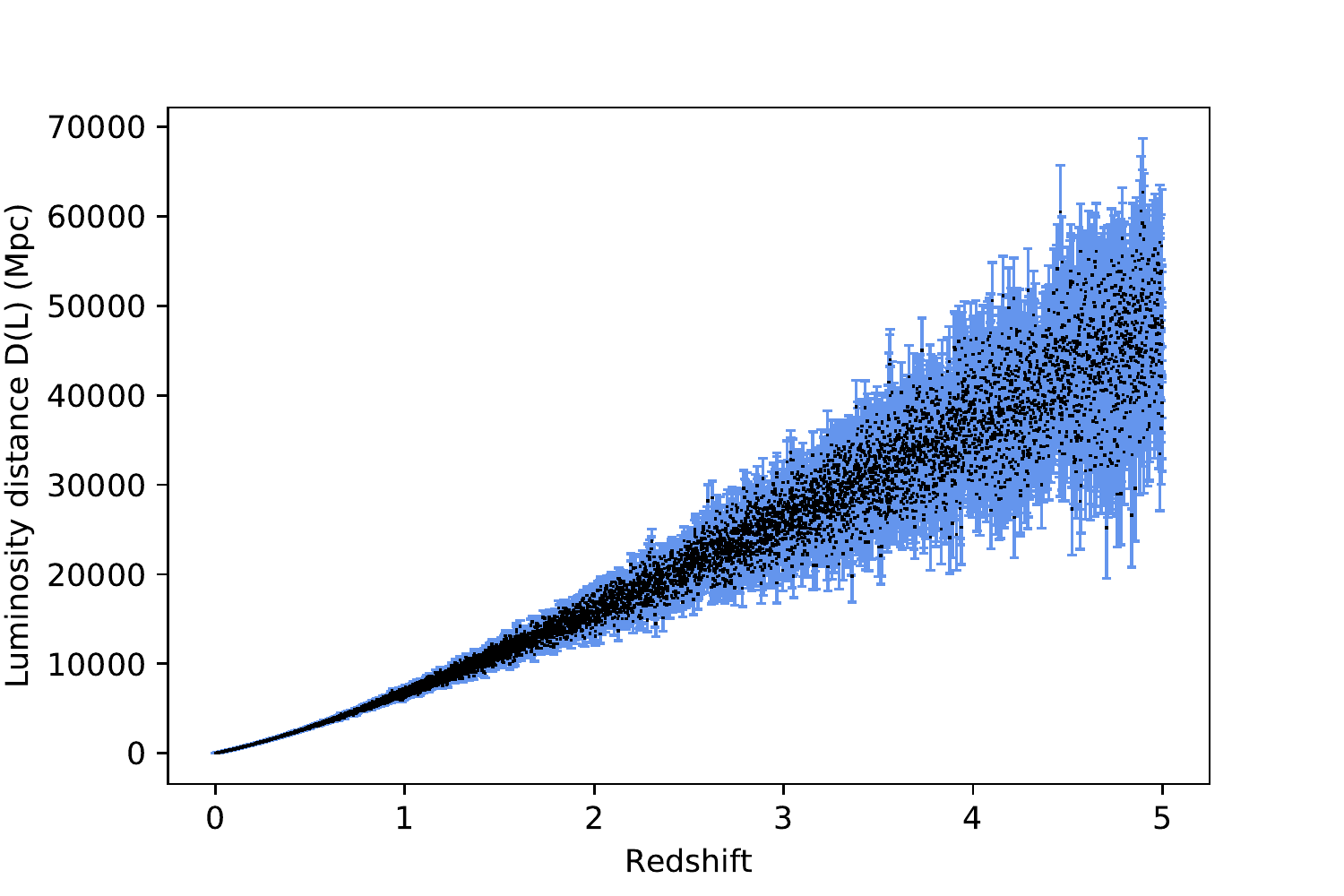}
\end{center}
\caption{The scatter plot of the luminosity distances in the catalog
of 10,000 simulated GW events detectable by the DECIGO.}
\end{figure}

\begin{figure}
\begin{center}
\includegraphics[width=0.95\linewidth]{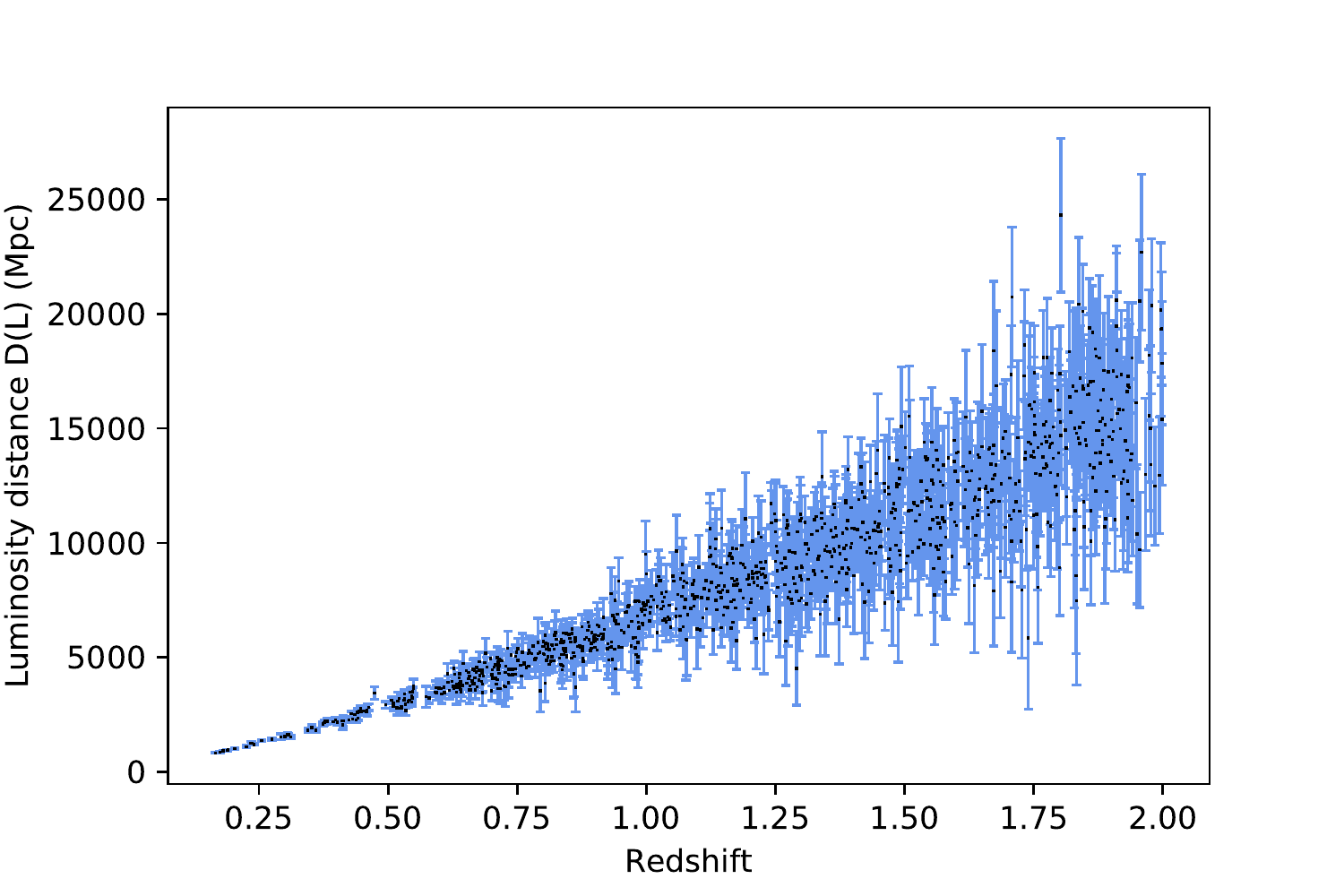}
\end{center}
\caption{The scatter plot of the luminosity distances in the catalog
of 1000 simulated GW events observable by the third generation
detector ET.}
\end{figure}

Our goal is to employ Eq.(\ref{eq3}) for the assessment of cosmic
curvature. For this purpose we need not only  $D_L(z)$ but also the
derivative of this function. Such derivative can be reconstructed by
many different methods. We have chosen the following approach.
First, we make an empirical fit to the luminosity distance based on
a third-order logarithmic polynomial as in \citet{Risaliti2018}
\begin{equation}\label{eq11}
D_L^{P}(z)=\frac{c}{H_0}ln(10)[x+a_2x^2+a_3x^3 + {\cal O}(x^4)]
\end{equation}
where $x = \log(1+z)$, $a_2$ and $a_3$ are two constants that need
to be fitted to the standard siren data sets. Compared to other
empirical fitting methods, such logarithmic parametrization has an
advantage of faster convergence at higher redshifts ($0<z<0$). Let us remark that cosmography based on the logarithmic polynomial expansion is expected to break down at $z > 2$ and produce spurious effects as discussed in \cite{Banerjee2020}. We will discuss it further in the next section. For now, let us stress that in the redshift range of $H(z)$ on which we fit our cosmographic formula \ref{eq11} the convergence is reasonable.  More
specifically, to obtain the best-fit values of parameters and their
uncertainties, we use the Markov Chain Monte Carlo (MCMC) method
introduced by \citet{Foreman-Mackey2013} and implemented in the
Python module called emcee
\footnote{https://pypi.python.org/pypi/emcee}. For the standard
siren data sets, the parameters $a_2$ and $a_3$ characterizing the
luminosity distance are optimized by minimizing the $\chi^2$
objective function
 \begin{equation}\label{eq12}
\chi^2=\sum \limits_{i=1}^{n}{\frac{[ D_L^{GW}-
D_L^{P}]^2}{\sigma_{GW}^2}},
\end{equation}
where $\sigma_{GW}^2$ can be obtained from the Eq.~(9). The
marginalized probability distribution of each parameter and the
corresponding marginalized 2-D confidence contours are presented in
Figs.~3-4.

\begin{figure}
\begin{center}
\includegraphics[width=0.95\linewidth]{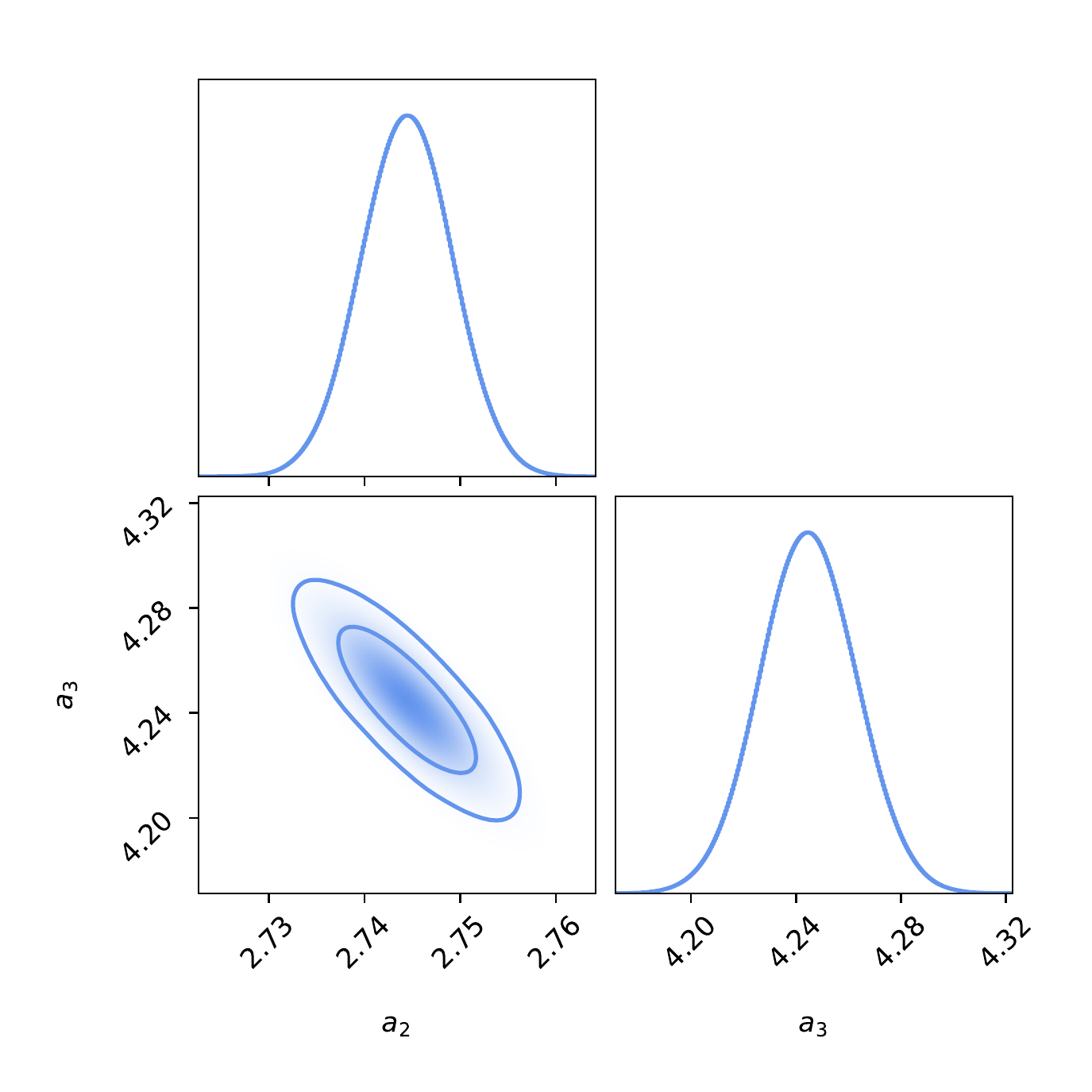}
\end{center}
\caption{Marginalized constraints on $a_2$ and $a_3$ based on the
standard sirens catalog from the DECIGO.}
\end{figure}

\begin{figure}
\begin{center}
\includegraphics[width=0.95\linewidth]{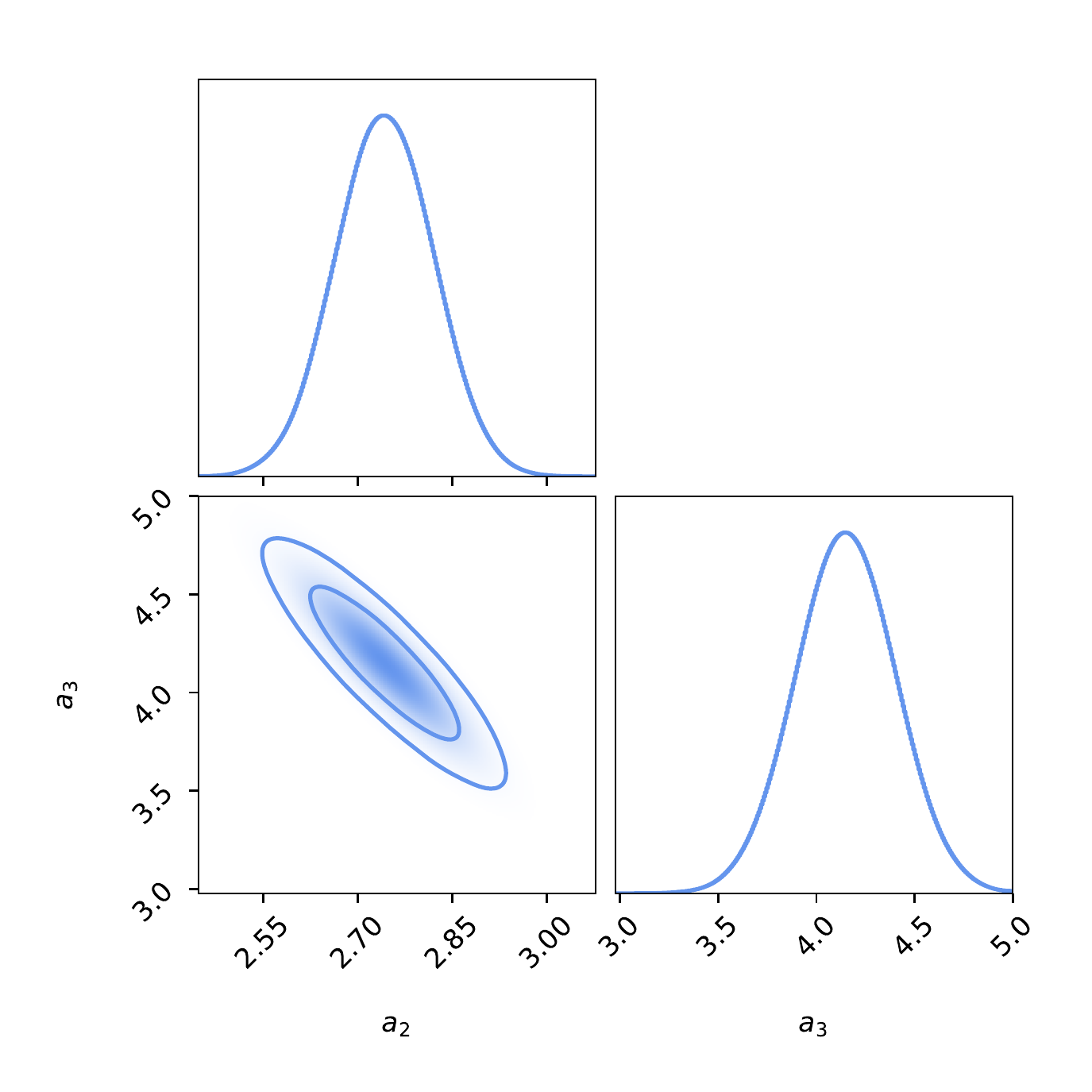}
\end{center}
\caption{Marginalized constraints on $a_2$ and $a_3$ based on the
standard sirens catalog from the third generation GW detector ET.}
\end{figure}

\subsection{The expansion rate measurements $H(z)$ }
The expansion rate (i.e., the Hubble parameter $H(z)$) measurements
have been widely used in the research of the nature of dark energy
and the evolution of the Universe. The idea of using the
differential age evolution of passively evolving galaxies has been
proposed by \citet{Jimenez2002} as a method independent on any
assumptions regarding the details of the cosmological model. Namely,
one can write the definition of the expansion rate in the following
form:
\begin{equation}\label{eq13}
H(z)=\frac{\dot{a}}{a}=-\frac{1}{1+z}\frac{dz}{dt}.
\end{equation}
The derivative of the redshift $z$ with respect to cosmic time $t$
can be directly obtained by measuring the age difference between the
two galaxies at different redshifts. In the case of early-type
galaxies evolving passively on a timescale longer than their age
difference, certain features of their spectra, such as the D4000
break enable us to measure the age difference of such galaxies. This
is so-called differential age (DA) or cosmic chronometer (CC)
approach. There is also another approach to obtain the $H(z)$ data
based on the radial baryon acoustic oscillations (BAO) features from
galaxy clustering
\citep{Gaztaaga2009,Blake2012,Samushia2013,Font-Ribera2014,Delubac2015}.
However, the $H(z)$ data obtained employing such method are based on
an assumed fiducial cosmological model and the prior for the
distance to the last scattering surface from CMB observations
\citep{Li2016c}. Moreover, there exist a deviation between
those two $H(z)$ compilations \citep{Zheng2016}. 
Therefore, we do not use $H(z)$ data obtained from
BAO. Consequently, we use only the latest CC  measurements of $H(z)$
comprising totally 31 data-points covering the redshift range
$0.070<z<1.965$.

\begin{figure}
\begin{center}
\includegraphics[width=0.95\linewidth]{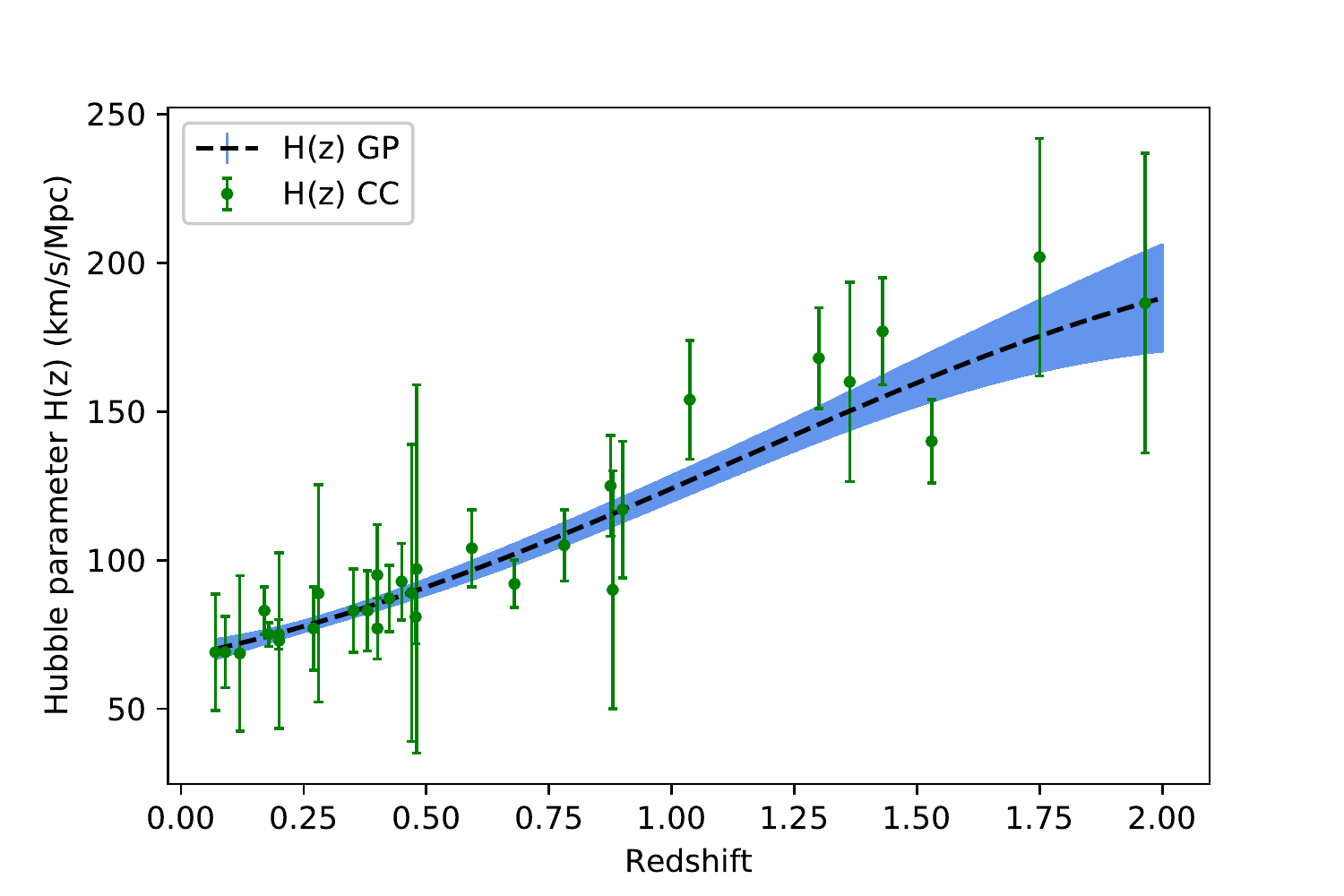}
\end{center}
\caption{The latest measurements of 31 Hubble parameters from the
galaxy differential age method and the $H(z)$ reconstruction
function with GP. Green dots denote CC measurements with
corresponding 1-$\sigma$ uncertainties. Black line and blue region
represent the reconstructed $H(z)$ function and its 1-$\sigma$
uncertainty band.}
\end{figure}

Besides using only a small sample of available $H(z)$ data to assess
the cosmic curvature we also reconstruct the $H(z)$ function from
the data. For this purpose we employ a smoothing technique based on
Gaussian processes (GP). Such an approach, firstly proposed by
\citep{Seikel2012a}, has been extensively applied in various studies
\citep{Shafieloo2012,Seikel2013,Yahya2014,Cai2016,Zheng2020}. The
crucial issue in the GP technique is to determine the covariance
function, with which one can derive the quantities at the redshifts,
where they have not been directly measured. In this work, we use the
squared exponential covariance function
\begin{equation}\label{eq15}
k(x,\tilde{x})=\sigma^2_{f}exp(-\frac{(x-\tilde{x})^2}{2\ell^2}),
\end{equation}
where $x$ and $\tilde{x}$ are any two different points, $\ell$ and
$\sigma_{f}$ are two hyperparameters, which characterize the
bumpiness of the function and can be optimized by GP with the
observed data set. $\ell$ can be thought of as the characteristic
length one has to travel in $x$-direction to get a significant
change in $f(x)$, whereas $\sigma_{f}$ denotes the typical change in
the y-direction. In contrast to the Mat\'{e}rn and Cauchy covariance
function, which are another popular choices, the squared exponential
covariance function has the advantage that it is infinitely
differentiable, which is useful for reconstructing the derivative of
the function \citep{Seikel2012a,Zheng2020}. By using the squared
exponential covariance function and the zero mean function  $\mu(z)$
(another ingredient of GP approach), we can obtain the values of
data points at other redshifts which have not be observed. This can
effectively bridge the redshift gaps between current data-points.
Technically, we use the code called GaPP in Python
\footnote{https://github.com/carlosandrepaes/GaPP} to derive our GP
results. The latest measurements of 31 Hubble parameters obtained
from CCs and the resulting GP reconstruction of $H(z)$ function are
presented in Fig.~5.

Now, we combine the reconstructed $D(z)$ and $D'(z)$ with the Hubble
parameter $H(z)$ to constrain the $\Omega_k(z)$ according to
Eq.~(3). Let us stress again that the standard sirens and cosmic
chronometers are independent of each other, which indicates that
such determination of the cosmic curvature is model-independent.

\section{Results and discussion}

By applying the above described procedure to $D_L(z)$ reconstructed
from simulated catalog of standard sirens observable by DECIGO and
$H(z)$ data from cosmic chronometers, we obtain 31 independent
measurements of the cosmic curvature $\Omega_k(z)$ at each redshift
$z$, where $H(z)$ is measured. They are shown in Fig.~6. The
uncertainty bars displayed on this figure were calculated from the
uncertainty budget of $D(z)$, $D'(z)$ and $H(z)$.

One can summarize these measurements with the inverse variance
weighted mean
\begin{equation}
\begin{array}{l}
{\bar \Omega}_k=\frac{\sum\left(\Omega_{k,i}/\sigma^2_{\Omega_{k,i}}\right)}{\sum1/\sigma^2_{\Omega_{k,i}}},\\
\sigma^2_{{\bar
\Omega}_{k}}=\frac{1}{\sum1/\sigma^2_{\Omega_{k,i}}},
\end{array}
\end{equation}
where ${\bar \Omega}_{k}$ stands for the weighted mean of cosmic
curvature and $\sigma_{{\bar \Omega}_{k}}$ is its uncertainty. We
find that, combination of future DECIGO standard sirens and
currently available CC data gives ${\bar \Omega_k}=0.004\pm0.09$. It
is fully consistent with the vanishing curvature -- an assumption
underlying our GW data simulations and compatible with the
constraints obtained from the latest Planck CMB measurements
\citep{Planck Collaboration2018}. Let us recall that we are
combining simulated and real data. The precision of this estimate is
comparable to the recent analysis of \citet{Cao2019b}, which
represented the precision of $\Delta\Omega_k\sim10^{-2}$ with 250
well-observed radio quasars. Such conclusion is also well consistent
with the cosmic curvature  derived from the strong lensing and
supernova distance measurements in the framework of another
model-independent test based on the distance sum rule
\citep{Zhou2020}.

\begin{figure}
\begin{center}
\includegraphics[width=0.95\linewidth]{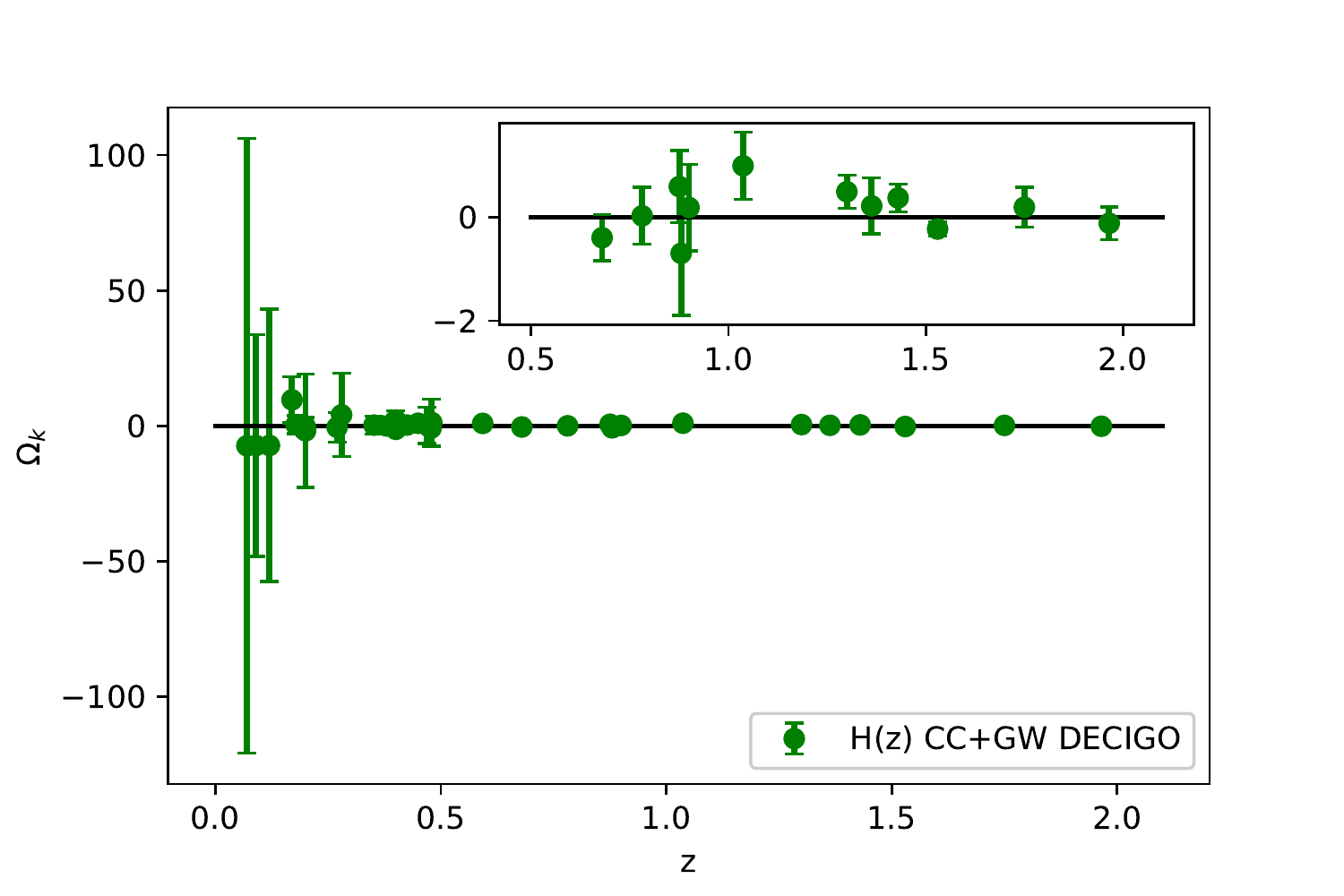}
\end{center}
\caption{31 measurements of the cosmic curvature parameter
$\Omega_k$ from the standard sirens of GW space-based detector
DECIGO and cosmic chronometers. Inset in the upper right reveals
more details. }
\end{figure}

\begin{figure}
\begin{center}
\includegraphics[width=0.95\linewidth]{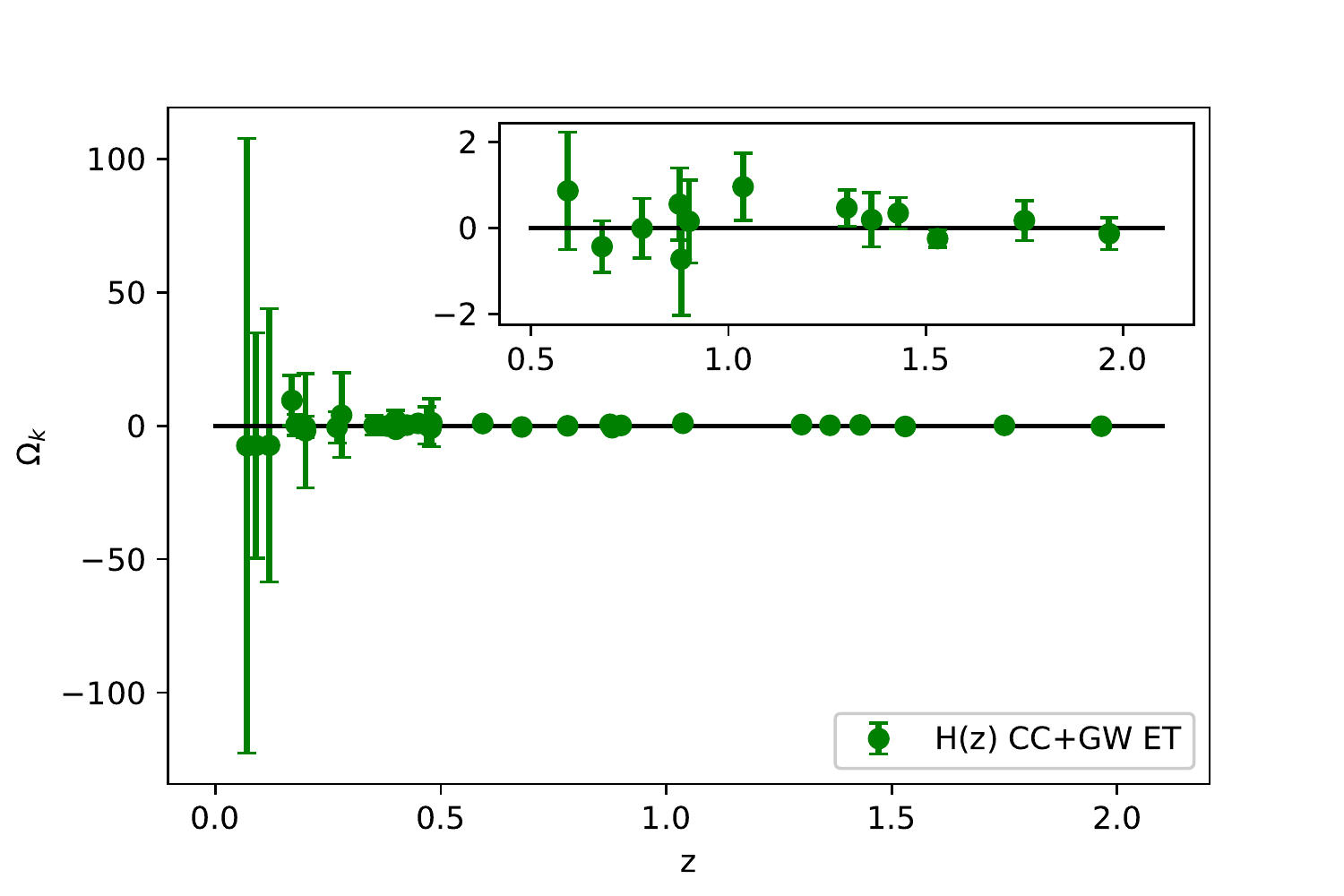}
\end{center}
\caption{31 measurements of the cosmic curvature parameter
$\Omega_k$ from the standard sirens of GW third-generation
ground-based detector ET and cosmic chronometers. Inset in the upper
right reveals more details. }
\end{figure}

\begin{table}
\begin{tabular}{c| c| c }
\hline\hline
  redshift bins & $CC+DECIGO$ & $CC+ET$  \\
\hline
 $0-2.0$    &  $0.004\pm0.09$     & $0.009\pm0.13$  \\
\hline
 $0-0.5$      & $-0.22\pm0.63$    & $-0.23\pm0.74$  \\
\hline
 $0.5-1.0$  &$-0.02\pm0.27$    & $-0.03\pm0.34$  \\
\hline
 $1.0-1.5$  &$0.45\pm0.19 $    & $0.41\pm0.24$\\
\hline
$1.5-2.0$   &$ -0.17\pm0.12$    & $-0.17\pm0.17$ \\
\hline\hline
\end{tabular}
\caption{Weighted average of $\Omega_k$ in different redshift bins
from the standard siren of GW detector (DECIGO and ET) and standard
clock of cosmic chronometers H(z).} \label{1}
\end{table}

When the current CC data are combined with forecasts for the ET
ground-based detector, one obtains ${\bar \Omega_k}= 0.009\pm0.13$.
Individual reconstructed $\Omega_k(z)$ are shown in Fig.~7. Again
the results are compatible with vanishing cosmic curvature, however
the precision of the final result is much worse.

Since our simulated GW catalogs used to derive $D_(z)$ and $D'_L(z)$
are rich enough we attempted at dividing the full sample into
different redshift bins and assessed ${\bar \Omega_k}$ in each bin.
More specific, we divide the $H(z)$ measurements into four groups,
including 18 data points with redshifts $z<0.5$, 6 data points with
redshifts $0.5<z<1.0$, 4 data points with redshifts $1.0<z<1.5$ and
3 data points with $1.5<z<2.0$. The cosmic curvature parameter
obtained in these sub-samples is presented in Table.~1. Note that
the derived curvature is negative when $z>1.5$, which is also
consistent with the results from the model-dependent constraints in
the literature \citep{Cai2016}.

\begin{table}
\begin{tabular}{c | c | c}
\hline\hline
  redshift bins & 4th-order polynomial & 5th-order polynomial  \\
\hline
$0-2.0$    & $0.01\pm0.10$    & $0.0001\pm0.11$  \\
\hline
$0-0.5$    & $-0.29\pm0.66$   & $-0.25\pm0.66$ \\
\hline
$0.5-1.0$  & $-0.06\pm0.29$   & $-0.07\pm0.28$ \\
\hline
$1.0-1.5$  & $0.46\pm0.20$    & $0.42\pm0.22$  \\
\hline
$1.5-2.0$  & $ -0.15\pm0.13$  & $-0.16\pm0.15$ \\
\hline\hline
\end{tabular}
\caption{Weighted average of $\Omega_k$ in different redshift bins
from CC+DECIGO based on fourth and fifth order logarithmic polynomial.} \label{table2}
\end{table}

Finally, as mentioned in  \citep{Banerjee2020} 
the third-order logarithmic polynomial cannot be trusted at high redshift range 
(especially $z>2$ region) and the influence of the breakdown of the validity of log-polynomial expansion was discussed in that paper. 
In order to check the 
performance of our approach based on the third order expansion and to see how it relates to the findings of \cite{Banerjee2020}, we repeated the whole procedure with 4th and 5th-order logarithmic polynomial expansions for different redshift bins.
Corresponding results are listed in Table\ref{table2}. It can be seen that next orders of expansion give the results compatible to the third order one. Of course, the order at which we truncate the logarithmic polynomial expansion affects the final result, but up to $z\approx 2$ the deviations are not dramatic, which is consistent with previous  investigations  \cite{Banerjee2020,Yang2020}. 

Moreover, we performed the $\Omega_k(z)$ reconstruction using not the
point (or binned) CC data but the $H(z)$ reconstruction using a
model-independent smoothing technique of Gaussian processes
\citep{Seikel2012a}. Fig.~8 displays such reconstructed $\Omega_k$
parameter as a function of redshift for two different scenarios:
combined CC + DECIGO data and CC + ET data. We expect that as the
precision of the future data improves (especially the Hubble
parameter $H(z)$ based on the full sky BAO survey up to redshift $z
\sim5$ \citep{Weinberg2013}), our approach will yield an even more
accurate determination of $\Omega_k$, especially at higher redshifts
\citep{Yu2016}.
We also repeated the $\Omega_k$ reconstruction with cosmography based on 
4th and 5th-order log polynomial expansions. The results turned out indistinguishable from  those shown in Fig.~8. Certainly, the impact of the breakdown of polynomial expansions should be investigated before any cosmological applications especially involving data at high redshifts. This would be particularly important in the precision cosmology era.

\section{Conclusions}

In this paper, we applied a model-independent technique to constrain
the cosmic curvature parameter $\Omega_k$ at different redshift
points directly, based on the idea expressed by
\citet{Clarkson2007}. Because currently we do not have samples of
standard candles ($D_L(z)$ measurements) rich enough to implement
this method, we used simulations of standard sirens data obtainable
from DECIGO and ET future space-borne and ground-based detectors.
They are expected to register ${\cal O}(10^4)$ NS-NS inspirals up to
$z\sim5$ or ${\cal O}(10^4)$ such events up to $z\sim2$,
respectively. The second necessary ingredient of this method are
independently obtained measurements of expansion rates at different
redshift. For this purpose we used the current data obtained from
passively evolving galaxies (CC) covering the redshift range of
$0.07<z<2$.

Firstly, from simulated $D_L$ samples representing future standard
siren observations (DECIGO and ET) we reconstructed the transverse
comoving distance $D(z)$ and its derivative $D'(z)$ with respect to
redshift $z$. In order to achieve this goal not relying on any
specific cosmological model, we used the third-order logarithm
polynomial approximation of $D_L(z)$ with undetermined coefficients,
which could be optimized numerically. Then, one was able to directly
calculate the curvature parameter $\Omega_k$ combining these results
with the expansion rate $H(z)$ measurements obtained from a sample
of cosmic chronometers.

\begin{figure}
\begin{center}
\includegraphics[width=0.9\linewidth]{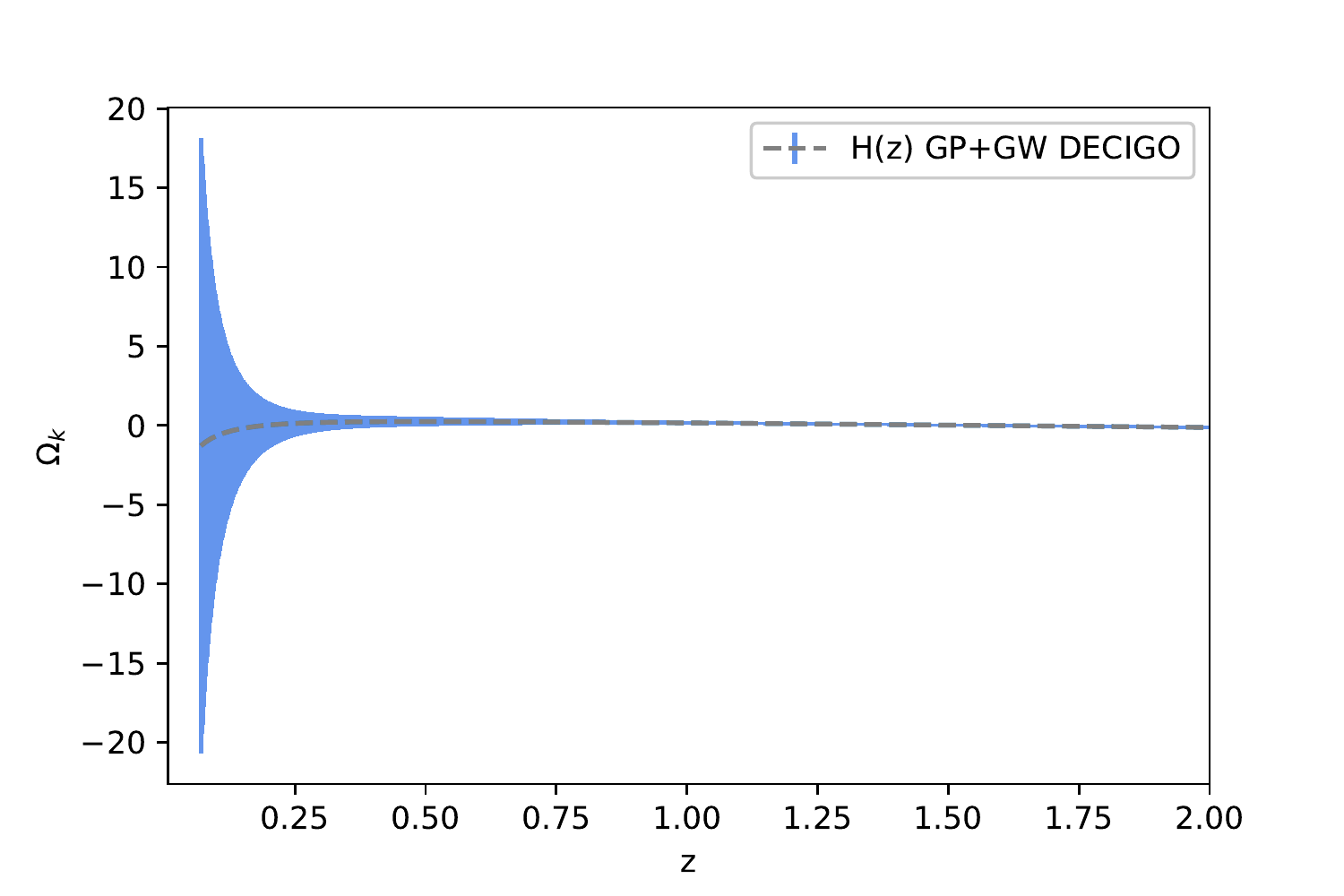}
\includegraphics[width=0.9\linewidth]{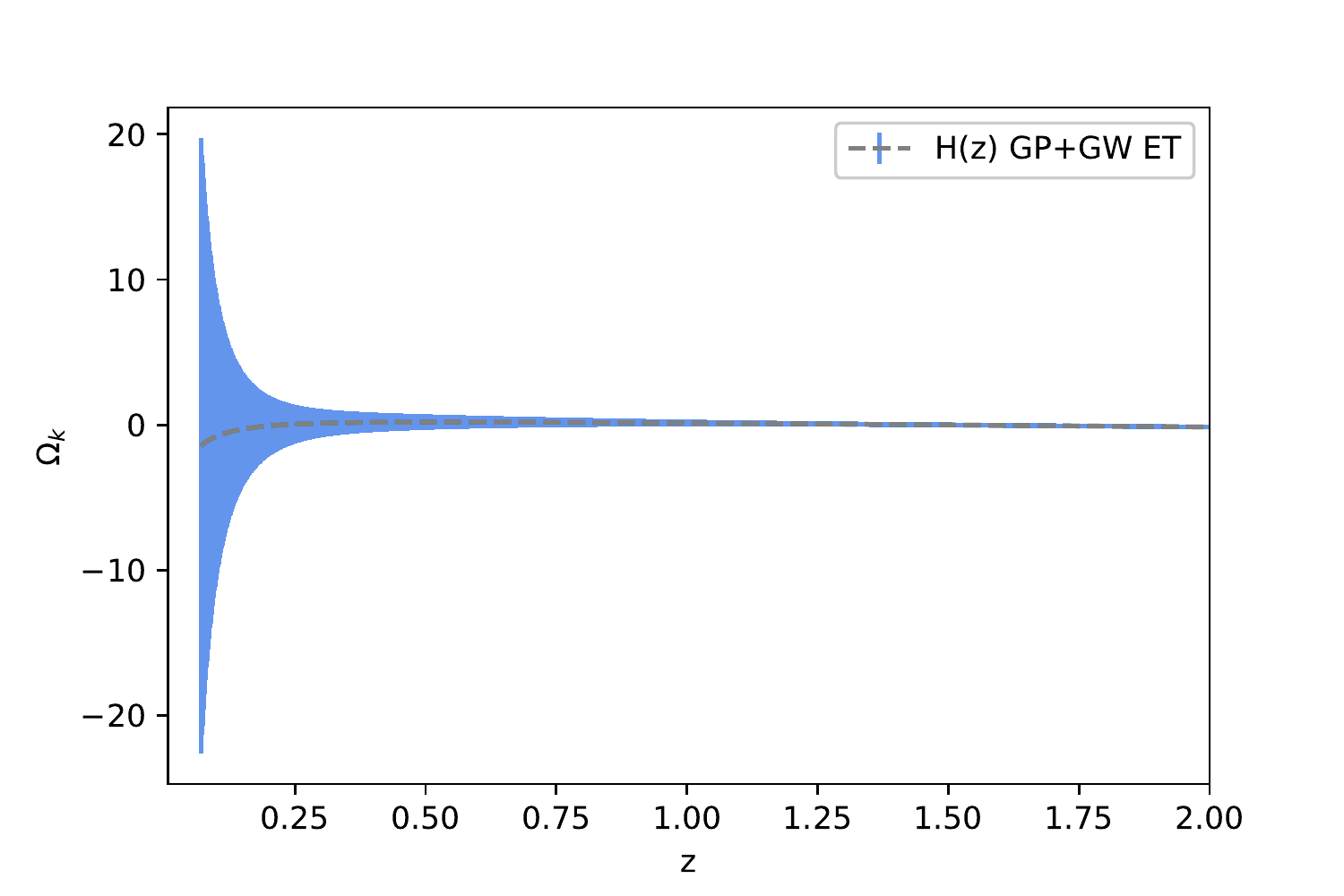}
\end{center}
\caption{Gaussian  process reconstruction of $\Omega_k$ obtained
from the combined data sets of $H(z)+DECIGO$ (upper) and $H(z)+ET$
(lower). The blue regions are the 68\% C.L. of the reconstructions.}
\end{figure}\label{fig:gpok}

We found that the cosmic curvature could be constrained as
$\Omega_k=0.004\pm0.09$ when the DECIGO detected standard sirens and
cosmic chronometers were used. When the standard siren (ET) is
considered, the cosmic curvature estimate is $\Omega_k=
0.009\pm0.13$. In other words the expected precision is $\Delta
\Omega_k = 10^{-2}$ with the DECIGO+CC and $\Delta \Omega_k =
10^{-1}$ with the ET+CC. Compared to the latest model-independent
estimations of the spatial curvature using the distance sum rule
method \citep{Zhou2020}, our results suggest a considerable
improvement in precision when the future GW observatories provide
rich statistics of $D_L(z)$ measurements. In order to surpass the
limitations of small CC sample we reconstructed $H(z)$ function
using GP technique and used it together with reconstructed $D_L(z)$
and $D'_L(z)$ to investigate $\Omega_k(z)$. This is important to
look if the current data, even though scarce, yet contain hints that
cosmic curvature could locally deviate from the overall curvature
assumed in FLRW metric. Such possibilities could be expected as a
result of back-reaction of inhomogeneities, which entered non-linear
regime during the structure formation \citep{Bolejko}.

Summarizing, the newly emerged gravitational wave astronomy can
acquire another dimension of being useful in local measurements of
cosmic curvature using distant sources. Such approach, essentially
different from doing fits of globally defined (at the level of FLRW
metric) curvature parameter. Possible deviations between these two
approaches might be an extremely useful hints of phenomena and
processes overlooked in current cosmological studies.

\section*{Acknowledgements}
We would like to thank Jingzhao Qi and Sixuan Zhang for their
helpful discussions. This work was supported by National Key R\&D
Program of China No. 2017YFA0402600; the National Natural Science
Foundation of China under Grants Nos. 12021003, 11690023, and
11633001; Beijing Talents Fund of Organization Department of Beijing
Municipal Committee of the CPC; the Strategic Priority Research
Program of the Chinese Academy of Sciences, Grant No. XDB23000000;
the Interdiscipline Research Funds of Beijing Normal University; and
the Opening Project of Key Laboratory of Computational Astrophysics,
National Astronomical Observatories, Chinese Academy of Sciences.
J.-Z. Qi was supported by China Postdoctoral Science Foundation
under Grant No. 2017M620661, and the Fundamental Research Funds for
the Central Universities N180503014. M.B. was supported by the
Foreign Talent Introducing Project and Special Fund Support of
Foreign Knowledge Introducing Project in China. He was supported by
the Key Foreign Expert Program for the Central Universities No.
X2018002.


\label{lastpage}

\maketitle

\end{document}